# Anisotropy-Driven Anomalous Inverse Orbital Hall Effect in Fe Films

E. Santos,[1] U. Borges,[1] J. L. Costa,[1] J. B. S. Mendes,[2] and A. Azevedo[1]

[1]*Departamento de Física, Universidade Federal de Pernambuco, 50670-901 Recife, Pernambuco, Brasil*
[2]*Departamento de Física, Universidade Federal de Viçosa, 36570-900 Viçosa, Minas Gerais, Brasil*

This study investigates the anomalous orbital effects in iron (Fe) films with strong uniaxial anisotropy, highlighting the interactions between spin and orbital currents. Using heterogeneous YIG/Fe and YIG/Pt/Fe structures, fabricated by oblique deposition in a magnetic field, spin pumping ferromagnetic resonance (SP-FMR) measurements were performed. It was observed that the uniaxial anisotropy enables the emergence of spin-to-charge (AISHE) and orbital-to-charge (AIOHE) conversion signals in out-of-plane configurations, where the spin polarization is parallel to the direction of the spin current. Experimental analysis revealed that orbital dynamics, mediated by orbital Hall conductivity, are more prominent in Fe films due to the low spin-orbit interaction (SOC) and high orbital response. These findings provide fundamental insights for the advancement of orbitronics devices, indicating the potential for controlling orbital and spin currents through magnetic and anisotropic parameters.

## I. INTRODUCTION

The discovery that it is possible to generate orbital angular momentum (OAM) flux in solids has led to the emergence of a new field of study focused on the transport of the orbital degree of freedom, now known as orbitronics.[1,2] Theoretical predictions[1] and experimental results[3] have shown that orbital currents can be created by applying a perpendicular electric field, known as the orbital Hall effect (OHE), which arises from nonequilibrium interband superpositions of Bloch states with different orbital characters.[1] In addition to the OHE, the inverse orbital Hall effect (IOHE) was discovered,[4] in which a charge current can be generated due to an orbital current, making it a highly robust effect that can serve as a probe for the study of various materials and interfaces.

One of the main advantages of the inverse orbital Hall effect (IOHE), over its spin analogue, is the ability to generate electrical signals of large magnitude, even in materials with weak spin-orbit coupling (SOC).[5] Materials such as titanium (Ti), ruthenium (Ru), germanium (Ge) and copper oxide (CuOx) have already been investigated, showing significant values for orbital-to-charge conversion.[4-7]

Spin effects naturally emerge in materials with strong spin-orbit coupling (SOC),[5] arising from the interaction between the lattice's orbital angular momentum and the electronic spin angular momentum (SAM). This mechanism underscores the central role of the orbital response in the dynamics of OAM in solids, suggesting that orbital transport plays a more fundamental role than spin transport.[1,8] Among the most significant spin effects are the spin Hall effect (SHE),[9–11] which generates spin currents from charge currents, and the inverse spin Hall effect (ISHE),[12,13] in which a spin current induces a perpendicular charge current. Conventionally, the charge current generated by the inverse spin Hall effect (ISHE) depends on the spin Hall conductivity ($\sigma_{SH}$), which is generally represented as a tensor. Consequently, the ISHE signals become more complex to analyze, as the interconversion between spin and charge involves an order parameter.[14] This effect is known as the anomalous spin Hall effect (ASHE), and its inverse effect (AISHE) has already been experimentally verified for spin currents in magnetic materials[15] and orbital currents in antiferromagnetic materials[16].

Experimental results[16] have demonstrated the existence of an orbital Hall conductivity tensor $\sigma_{OH}$, whose behavior is similar to that of the spin Hall conductivity tensor $\sigma_{SH}$, but with a significantly larger magnitude. Theoretically, the magnitudes of the components of the conventional spin Hall (SH) and orbital Hall (OH) conductivities $\sigma_{SH,OH}$ have been calculated using various methods[17,18] Magnetic iron (Fe) is notable for its strong orbital Hall conductivity $\sigma_{OH} \sim 2345\,(\hbar/e)(\Omega \cdot cm)^{-1}$, and relatively small spin Hall conductivity $\sigma_{SH} \sim 587\,(\hbar/e)(\Omega \cdot cm)^{-1}$. As such, it is a promising candidate for studying the effects of orbital-to-charge interconversion through both the IOHE and the anomalous IOHE.

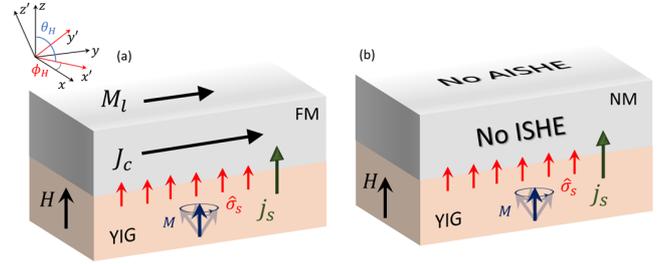

**Figure 1**. A spin current with spin flow and polarization $\hat{\sigma}_s$ in the z-direction is generated in a YIG film and pumped into an adjacent NM (a) or FM (b) layer. When the top layer is FM, spin-to-charge conversion occurs due to the AISHE. However, when the top layer is NM, no charge current is detected.

In this context, our study presents a systematic investigation of FM/Fe(12) and FM/Pt($t_{Pt}$)/Fe(12) heterostructures, with $t_{Pt} = 0, 2, 4, 6, 8, 10$ nm, where FM consists of yttrium iron garnet (YIG) ferrimagnetic insulator substrates fabricated using the liquid-phase epitaxy (LPE) technique. The substrates were cut into identical dimensions of 3.0×1.5 mm². The deposition of Pt and Fe thin films was carried out using DC sputtering at a base pressure of 1 mTorr.

The theoretical analysis of the SP-FMR signals was based on the effects of the inverse SHE and OHE, as well as the inverse anomalous SHE and OHE, using the values of $\sigma_{SH,OH}$ found in.[14] Experimental measurements were performed using the spin pumping driven by ferromagnetic resonance (SP-FMR) technique, with an RF power of 5 mW and a RF frequency of 9.51 GHz.

## II. EXPERIMENTAL RESULTS AND THEORETICAL DISCUSSION

Fig. 2(a) illustrates the SP-FMR process in a YIG/normal metal (NM) bilayer. During magnetization precession, a spin current density $j_s$ is injected into the NM layer, as described by the equation $j_s = \hbar g_{eff}^{\uparrow\downarrow}/4\pi M^2 (\vec{M} \times \dot{\vec{M}})$, as presented in.[19] This spin current consists of two components: an alternating current (AC), represented by the black vectors, and a direct current (DC), indicated by the fixed blue arrow, aligned parallel to the applied field. The DC component generates a spin accumulation that diffuses through the NM layer. This accumulation is characterized by spin polarization $\hat{\sigma}_s$, oriented along the z-axis, with a typical diffusion length of a few nanometers, particularly in materials with strong spin-orbit coupling (SOC).[5]

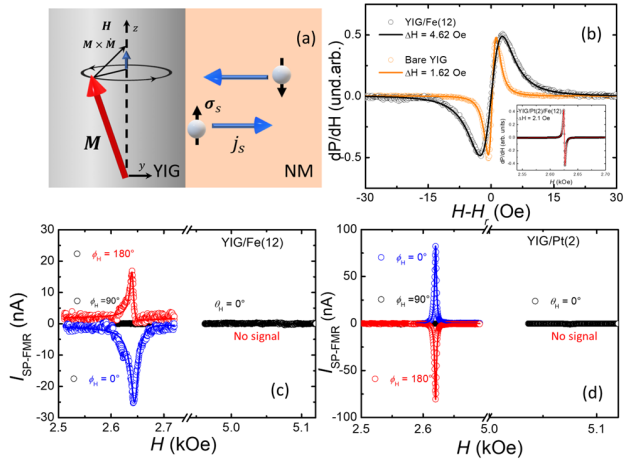

**Figure 2.** (a) Physical scheme of the spin pumping mechanism in bilayers. Under ferromagnetic resonance condition, the DC component $\vec{J}_S \propto (M \times \dot{M})$ accumulates at the interface and diffuses as a spin current with spin polarization $\hat{\sigma}_s$. (b) FMR curves for YIG (orange symbols) and YIG/Fe(12) (black symbols). The theoretical fits are the Lorentzian shape, where the line widths were obtained $\Delta H_{YIG} = 1.62$ Oe, and $\Delta H_{YIG/Fe} = 4.62$ Oe. The inset shows the result for YIG/Pt(2)/Fe(12), where $\Delta H_{YIG/Pt/Fe} = 2.1$ Oe, result of the decoupling between the YIG and Fe layers. (c) SP-FMR experimental data for YIG/Fe(12) for in-plane configuration ($H_r \sim 2.6$ kOe) and out-of-plane configuration ($H_r \sim 5$ kOe). For $\phi_H = 0°$, we have $I_{SP-FMR} < 0$ (blue symbols), for $\phi_H = 90°$, $I_{SP-FMR} = 0$ (black symbols), and for $\phi_H = 180°$, $I_{SP-FMR} > 0$ (red symbols). (d) SP-FMR experimental data for YIG/Pt(2), exhibiting the same behavior as (c) but with opposite SP-FMR signal polarities.

Fig. 2(b) shows the FMR curves for pure YIG (orange symbols) and for YIG/Fe(12) (black symbols). The solid curves are the experimental fits using Lorentzian shape, where we obtained linewidths $\Delta H_{YIG} = 1.62$ Oe and $\Delta H_{YIG/Fe} = 4.62$ Oe. The inset of Fig. 2(b) shows the FMR curve for YIG/Pt(2)/Fe(12), where we found $\Delta H_{YIG/Pt/Fe} = 2.1$ Oe. For the YIG/Fe(12) sample there was an increase of 3.0 Oe in the linewidth, while for the YIG/Pt(2) sample there was an increase of only 0.48 Oe, which we attribute to the additional damping in the magnetization dynamics equation due to the spin pumping process.[20] For the YIG/Fe(12) sample, in addition to the intrinsic and extrinsic mechanisms such as spin pumping, the exchange interaction between the spins of the magnetic layers results in an additional mechanism for the increase in the FMR linewidth.[21]

Fig. 2(c) and (d) show the SP-FMR signals for the YIG/Fe(12) and YIG/Pt(2) samples, respectively, where the electrical contacts were fixed with silver paint along the long axis of the sample. We observe a signal exhibiting negative polarity at $\phi_H = 0°$, positive polarity at $\phi_H = 180°$, and zero at $\phi_H = 90°$, assuming $\phi_H \cong \phi_M$. The magnitude of the charge current, given by $I_{SP-FMR} = V_{SP-FMR}/R$, at $\phi = 0°$, is 3.28 times larger in YIG/Pt(2), compared to the SP-FMR in YIG/Fe(12). This difference is attributed to the weak SOC of the Fe and its small $\sigma_{SH}$, consistent with Refs.[17,18] Furthermore, the experimental data indicate that $\sigma_{SH}^{Fe} < 0$, while $\sigma_{SH}^{Pt} > 0$, differing from the theoretical predictions if Refs.,[17,18] but aligning with the findings reported in Ref.[20] In the out-of-plane configuration ($\theta_H = 0°$), no signal was detected in the SP-FMR measurement. The results in the configuration in the plane obey the equation

$$J_c^k = (2e/\hbar)\theta_{ij}^{SH,OH} j_{s,L}^{ij}, \quad (1)$$

where, $\theta_{ij}^{SH,OH} = (e/\hbar)\sigma_{ij}^{SH,OH}/\sigma_e$ is a dimensionless coefficient known as the spin Hall angle ($\theta_{ij}^{SH}$), associated at spin Hall conductivity ($\sigma_{ij}^{SH}$), and orbital Hall angle ($\theta_{ij}^{OH}$), associated at orbital Hall conductivity ($\sigma_{ij}^{OH}$). The tensor $j_{s,L}^{ij}$ represents the spin or orbital current. In general, $\theta_{ij}^{SH,OH}$ is given by[14,16]

$$\theta_{ij}^{SH,OH} = \theta_0 \epsilon_{ijk} + [(\theta_1 + \theta_2)\delta_{ij}\delta_{kl\neq i} + \theta_1 \delta_{ik}\delta_{jl\neq i} + \theta_2 \delta_{il}\delta_{jk\neq i}]M_l, \quad (2)$$

The coefficients $\theta_n$ with $n = 0,1,2$ can be associated with spin or orbital effects, where the indices $i, j, k, l = 1,2,3$ correspond to the cartesian directions $x, y, z$. The index $i$ represents the direction of the spin/orbital current, $j$ indicates the polarization of the spin/orbital current, and $k$ refers to the direction of the charge current $J_c^k$. The first term in equation (2) represents the conventional ISHE/IOHE, while the additional terms are due to the AISHE or inverse anomalous orbital Hall effect (AIOHE).

Conventionally, when there is no AISHE/AIOHE, the experimental measurements are conducted in the film plane (for example, $k = 2$), where the spin polarization $\hat{\sigma}_s$ or orbital polarization $\hat{\sigma}_L$ also lies in the film plane, perpendicular to the direction of the experimental measurement (denoted by $j$), and the spin/orbital current flows perpendicular to the film plane ($i = 3$). Therefore, using these indices in equations (1) and (2), we obtain

$$J_c^2 = (2e/\hbar)\theta_0 \epsilon_{3j2} j_{s,L}^{3j}. \quad (3)$$

In vector notation, equation (3) becomes

$$J_c^2 = \theta_0 j_{s,L}^{3j}\left(\frac{2e}{\hbar}\right)\begin{bmatrix}1\\0\end{bmatrix}, \quad (4)$$

where the first element of the column vector refers to $j = 1$, and the second to $j = 2$. For a rotation of the axes by an angle $\phi_H$ with respect to the $x$-axis (Fig. 1(a)), the vector $J_c^{2'}$ in the new coordinate system is given by

$$J_c^{2'} = \begin{bmatrix}\cos\phi_H & -\sin\phi_H\\ \sin\phi_H & \cos\phi_H\end{bmatrix} J_c^2 = \theta_0 j_{s,L}^{3j}\left(\frac{2e}{\hbar}\right)\begin{bmatrix}\cos\phi_H\\ \sin\phi_H\end{bmatrix}. \quad (5)$$

The axis that coincides with the field $\vec{H}$ (or with the spin polarization $\hat{\sigma}_s$) is the $x'$-axis, where $j = 1$. Thus, the charge current $J_c^{2'}$ becomes,

$$J_c^{2'} = (2e/\hbar)\theta_0 j_{s,L}\cos\phi_H, \qquad (6)$$

which models the behavior of the ISHE or IOHE in the film plane. Note that, for $\phi_H = 0°$ the sign is positive, for $\phi_H = 90°$ the sign is zero, and for $\phi_H = 180°$ the sign is negative. For out-of-plane SP-FMR measurements varying only the polar angle $\theta_H$, using $i = 3, j, k = 2$, we obtain

$$J_c^{2'} = (2e/\hbar)\theta_0 j_{s,L}\sin\theta_H, \qquad (7)$$

which models the out-of-plane ISHE/IOHE behavior. In this case, for $\theta_H = 0°$ or $\theta_H = 180°$ the SP-FMR signal is zero. The physical origin of IOHE is related to the emergence of orbital textures in the crystal lattice, where the interaction with orbital currents creates an imbalance in the local OAM, responsible for the emergence of an orbital current. On the other hand, the spin current arises as a consequence of the SOC, which will couple the OAM and the SAM, simultaneously generating a coupled spin current and an orbital current, which can be converted into charge currents by distinct channels (IOHE or ISHE).

Unlike ISHE or IOHE, AISHE/AIOHE depend on a magnetic order parameter to exist, where this order parameter can be the Neél vector $\vec{n}_l$, for antiferromagnetic materials,[14,16] or the magnetization $M_l$, for ferromagnetic materials.[14,15]

Recent advances have shown that Berry curvature plays a central role in understanding spin and orbital transport phenomena, particularly their anomalous counterparts. The intrinsic Berry curvature mechanism, originally formulated to explain the anomalous Hall effect (AHE) in ferromagnetic materials such as (Ga,Mn)As and transition metals, has been extended to explain the SHE, ISHE, and now, their orbital analogs, including the OHE, IOHE, and their anomalous forms ASHE/AISHE and AOHE/AIOHE. These effects arise from the nonzero Berry curvature in the band structure, which acts as an effective magnetic field in momentum space, deflecting carriers transversely and generating measurable spin and orbital currents.[23–26]

Specifically, for materials with order parameters, the injection of spin/orbital currents is converted into charge current even in atypical experimental setups. For example, in the SP-FMR measurement, even if the spin/orbital polarization $\hat{\sigma}_{s,L}$ is in the same direction as the spin/orbital current, it is possible to convert it into charge current due to AISHE/AIOHE. In this setup, the spin/orbital current has polarization and direction along the z-direction, therefore

$$\theta_{33}^{SH,OH} = (\theta_1 + \theta_2)M_2, \qquad (8)$$

$$J_c^2 = (2e/\hbar)(\theta_1 + \theta_2)M_2 j_{s,L}^{33}, \qquad (9)$$

which is the equation for the AISHE/AIOHE signal represented in Fig. 1.

To investigate the anomalous signals in our samples it is necessary to ensure the existence of a magnetic order parameter. To verify this, we fabricated new YIG/Fe(12) samples with oblique DC sputtering deposition[27] coupled with an external magnetic field with a magnitude of 500 Oe applied along the major axis of the samples (inset Fig. 3(a)), which ensures the formation of a strong uniaxial anisotropy.

Fig. 3(a) shows the FMR signals for the YIG/Fe(12) sample, at $\phi_H = 0°$, $\theta_H = 90°$ at the Fe and YIG resonance, with Fe deposited in-plane ($\alpha = 0°$) and oblique ($\alpha = 60°$), where the in-plane angle $\phi_H$, and $\theta_H$ is defined as shown in the inset of Fig. 3(a). For the film of Fe deposited at $\alpha = 0°$ we find a resonance field $H_r = 630.0$ Oe, and $\Delta H = 83.0$ Oe. For $\alpha = 60°$ (Fe$^{60}$), $H_r = 1180.0$ Oe, and $\Delta H = 199.0$ Oe. The increase in the linewidth of Fe$^{60}$ compared to Fe is caused by variations in the internal magnetic field that can be seen as an extrinsic relaxation mechanism.[28] Fig. 3(b) shows the dependence of the in-plane magnetic field angle $\phi_H$ of the dP/dH FMR signal for the YIG/Fe(12) (black signals) and YIG/Fe$^{60}$(12) (green signals) samples, where we can clearly observe a strong uniaxial anisotropy, characterized by the strong symmetry breaking of the crystal lattice in the $\phi_H = 90°$ or $\phi_H = 270°$ direction. We can write the uniaxial magnetocrystalline surface energy density as follows[29] $E_u(\phi) = K_u \sin^2(\phi - \phi_u)$, where $\phi_u$ defines the direction of the uniaxial axis with respect to the crystal axes and $K_u$ is the first-order uniaxial anisotropy constant, respectively. The experimental fit of the data in Fig. 3(b) (black solid curve) was obtained using the Zeeman and uniaxial energy terms. The values of the parameters important for the fits are found in Appendix A.

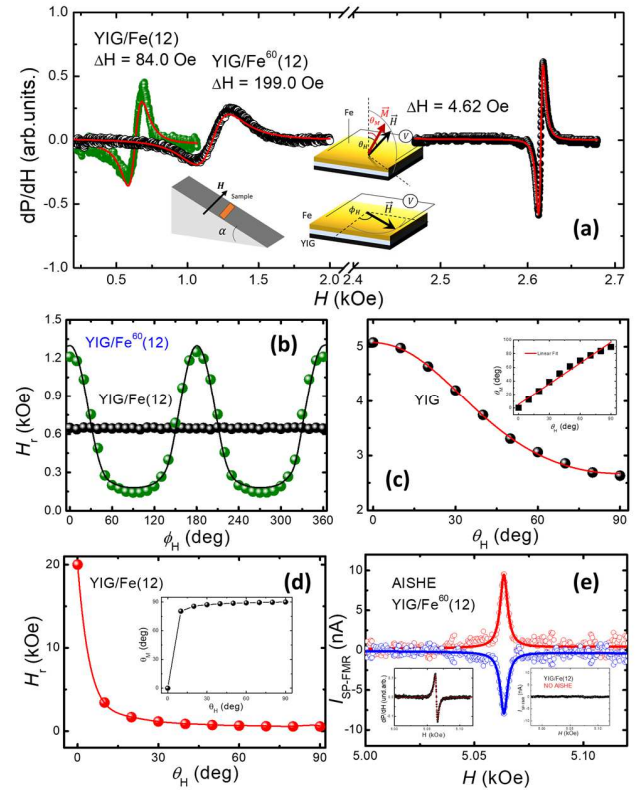

**Figure 3.** (a) FMR absorption curves for YIG/Fe(12) (green data), YIG/Fe$^{60}$(12) (black data), and YIG/Pt(2) (black data at $H_r$~2.6 kOe). The insets represent the angular configurations of the oblique sputtering deposition (angle $\alpha$) and the other insets represent the azimuthal $\phi_H$ and polar $\theta_H$ angles. (b) $H_r$ resonance field as a function of the angle $\phi_H$ ($0° \leq \phi_H \leq 360°$). For YIG/Fe(12) with deposition angle $\alpha = 0°$, no anisotropy is observed. On the other hand, when $\alpha = 60°$, a strong anisotropy appears, characterized by the easy axis at $\phi_H = 90°$ and $\phi_H = 270°$. (c) YIG resonance field data as a function of angle $\theta_H$ ($0° \leq \theta_H \leq 90°$). The inset shows the linear behavior between $\theta_H$ and $\theta_M^{YIG}$. (d) Fe resonance field data as a function of angle $\theta_H$. The inset shows the nonlinear behavior between $\theta_H$ and $\theta_M^{Fe}$, due to the strong demagnetization of Fe films. (e) AISHE signal for YIG/Fe$^{60}$(12).

Fig. 3(c) shows the dependence of the FMR signal dP/dH on the polar angle ($0° \leq \theta_H \leq 90°$). For the YIG sample $H_r$ was maximum for $\theta_H = 0°$ ($H_r \sim 5$ kOe), and minimum for $\theta_H = 90°$ ($H_r \sim 2.6$ kOe), due to the demagnetizing field, where. The fit (red solid curve) was obtained using the Zeeman energy term and the demagnetization energy term. The inset of Fig. 3(c) shows the linear behavior between the polar angle $\theta_H$ and $\theta_M$, which ensures that the magnetization of the YIG film aligns with the external magnetic field $\vec{H}$. Similarly, Fig. 3(d) shows the behavior of dP/dH as a function of ($10° \leq \theta_H \leq 90°$) for the YIG/Fe(12) sample at the Fe resonance. By extrapolating the theoretical fit curve, the best fit provides a resonance magnetic field of the order of 20 kOe for $\theta_H = 0°$, which is in agreement with the strong demagnetization energy for Fe films. The inset shows the relationship between $\theta_H$ and $\theta_M$ for Fe film, where we observe an abrupt change in magnetization at $\theta_H = 0°$, which shows that the magnetization $\vec{M}_{Fe}$ does not follow the magnetic field $\vec{H}$.

Based on the results presented in Fig. 3(d), it can be inferred that within the magnetic field range used in the SP-FMR measurements ($2.6 \leq H \leq 5.1$ kOe), the magnetization of the Fe remains confined to the plane of the film. Therefore, the YIG/Fe$^{60}$(12) sample has a well-defined order parameter, confined in the plane and with a strong uniaxial anisotropy along the major axis ($l = 2$).

Fig. 3(e) shows the SP-FMR signal for the YIG/Fe$^{60}$(12) sample, where the electrical contacts were fixed along the major axis of the sample, collinear with the easy axis of the Fe$^{60}$ film (order parameter). In the YIG resonance condition, the magnetization precession creates a spin accumulation at the YIG/Fe$^{60}$ interface that diffuses along the Fe$^{60}$ layer as a spin current and is converted to a charge current due to AISHE. Conventional ISHE cannot explain the observed signal. For $\theta_H = 0°$ (blue data) we find a signal with negative polarity with respect to YIG/Pt, and for $\theta_H = 180°$ we find a signal with positive polarity. This polarity can be inverted if the order parameter $M_2$ reverses direction $+y \rightarrow -y$ (equation 9). The signal inversion for $\theta_H = 180°$ is in agreement with equation 9 considering only the spin counterpart, since $j_s^{33} \rightarrow -j_s^{33}$. The insets of Fig. 3(e) show the FMR of YIG/Fe$^{60}$(12) at the YIG resonance that is in agreement with the SP-FMR signal, and the other inset shows the zero AISHE signal for the YIG/Fe(12) sample, due to the absence of an order parameter ($M_2 = 0$). For this sample we are not considering AIOHE, since YIG injects only spin current and Fe has weak SOC, as shown in Fig. 2(c).

The obtained AISHE signal has a considerable magnitude, around 10 nA, at a very low RF power of 5 mW, while the ISHE of YIG/Fe(12) was around 25 nA. Recent experimental results[3,5,16,17] show that the IOHE signals are usually larger than the ISHE signals due to $\sigma_{OH}$ being larger than $\sigma_{SH}$ in many cases. In ref.[16] it was observed that the AIOHE was larger than the AISHE. This motivated us to study the orbital dynamics in Fe films with strong uniaxial anisotropy from the IOHE and AIOHE. For this purpose, we fabricated new samples of YIG/Pt(t)/Fe(12) without anisotropy, to study the IOHE, and to study the AIOHE, we fabricated samples of YIG/Pt(2) and subsequently deposited Fe(12) films using oblique sputtering for $\alpha = 30°$, and oblique sputtering added to a magnetic field of 500 Oe for $\alpha = 60°$.

Fig. 4 shows the IOHE results on the YIG/Pt($t_{Pt}$)/Fe(12) samples without anisotropy. The films with anisotropy presented identical in-plane results, which is in agreement with equation (2). Fig. 4(a) shows the comparison between the SP-FMR signals for YIG/Pt(2) (red symbols) and YIG/Pt(2)/Fe(12) (blue symbols), where we observe a significant gain when adding the Fe(12) layer over the YIG/Pt(2) bilayer. Fig. 4(b) shows the SP-FMR signal only of YIG/Pt(2)/Fe(12) for the angles $\phi_H = 0°, 90°, 180°$ which obeys equation (6), and the zero signal for $\theta_H = 0°$, which is in agreement with equation (7), since Pt(2) has no order parameter. Fig. 4(c), (d), (e) and (f) show the SP-FMR signals for YIG/Pt($t_{Pt}$)/Fe(12), with $t_{Pt} = 4, 6, 8, 10$ nm, respectively. We notice a gradual reduction of the SP-FMR signal when increasing the thickness of the Pt film. To understand this reduction, we can base ourselves on the results of ref.[5], where they also observed a gradual reduction of the SP-FMR in YIG/Pt(2)/CuO$_x$(3) whose main effect is the inverse orbital Rashba-Edelstein effect (IOREE). The physical explanation for this is based on the strong SOC of Pt. By injecting a spin current into Pt films, coupling occurs between the spin and orbital states, resulting in a simultaneous spin-orbital flux that diffuses inside the Pt according to a diffusion equation, as described in ref.[30] It is therefore natural for relaxation of the spin states within Pt to occur, leading to a simultaneous relaxation of the orbital states. This behavior is illustrated in Fig. 4(g). For $t_{Pt} = 2$ nm, the SP-FMR signal exhibited the most significant enhancement. However, as $t_{Pt}$ increases, the signal tends to saturate, as observed at $t_{Pt} = 8$ nm and $t_{Pt} = 10$ nm. The inset of Fig. 4(g) shows the diffusion process of the spin-orbital current within the Pt exhibits $\lambda_{LS} = 1.1$ nm.

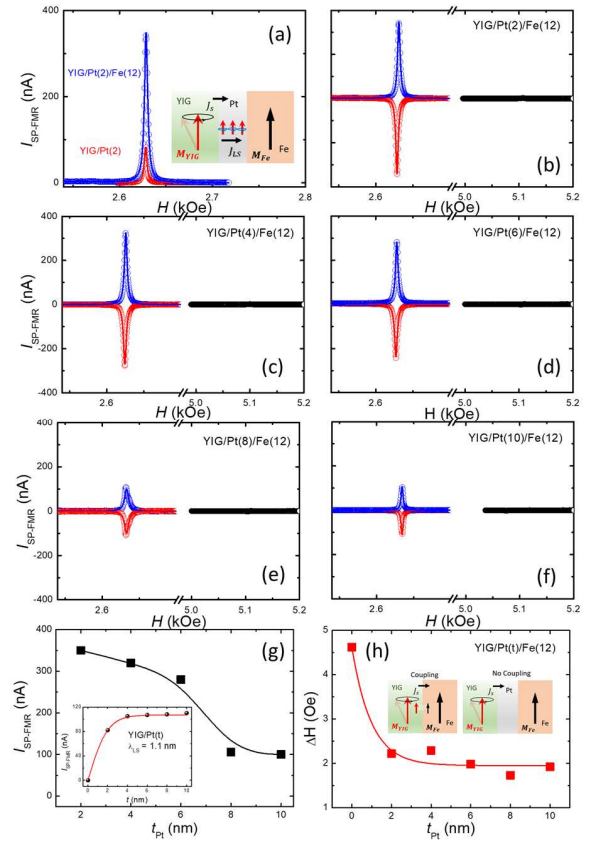

**Figure 4.** (a) Comparison between the SP-FMR signals of YIG/Pt(2) (red symbols) and YIG/Pt(2)/Fe(12) (blue symbols). The inset shows the physical process of the LS coupling within Pt. Figs (b),(c),(d), and (f) show the SP-FMR signals for YIG/Pt(t)/Fe(12). (g) Variation of the SP-FMR current $I_{SP-FMR}$ as a function of the Pt film thickness. (h) FMR linewidth of the YIG/Pt(t)/Fe(12). The inset illustrates the decoupling between the YIG and Fe layers upon adding an intermediate Pt layer.

Fig. 4(h) illustrates the variation in FMR linewidth ΔH for the YIG/Pt(t)/Fe(12) samples. A significant reduction in linewidth was observed upon introducing the Pt layer between YIG and Fe. This reduction can be attributed to the decoupling of the magnetics layers. For t≥2, the linewidth values showed only minor variations.

From the data in Fig. 4, the ratio between the SP-FMR signal measured for YIG/Pt(2)/Fe(12) and YIG/Pt(2) is 3.5 times. The signal from the first sample is a combination of the ISHE in Pt(2) and the IOHE in Fe(12), resulting in a signal of 350 nA (here, we neglect the ISHE from Fe). Subtracting the ISHE from YIG/Pt(2), a signal of approximately 250 nA remains, which is about 10 times larger than the ISHE in YIG/Fe(12), Fig. 1(c).

Finally, we performed measurements in the out-of-plane configuration on the YIG/Pt(2)/Fe(12), YIG/Pt(2)/Fe$^{30}$(12), and YIG/Pt(2)/Fe$^{60}$(12) samples. The results are shown in Fig. 5.

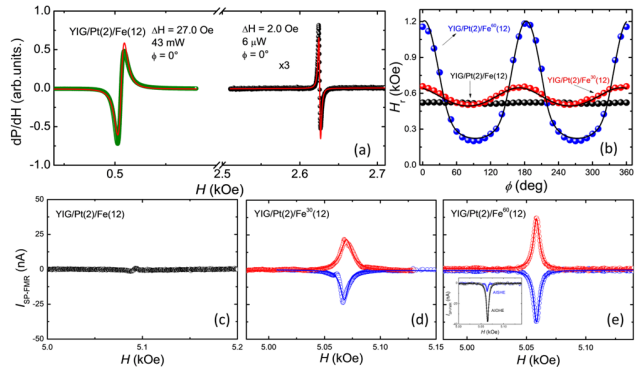

**Figure 5.** FMR measurements for YIG/Pt(2)/Fe(12). (b) SP-FMR measurements as a function of angle in the $\phi_H$ plane for three samples, YIG/Pt(2)/Fe(12) with $\alpha = 0°, 30°, 60°$, where the third sample presented strong uniaxial anisotropy. (c), (d) and (f) show the out-of-plane SP-FMR signals due to AISHE and AIOHE. The sample without anisotropy did not show any anomalous signal. The inset in (e) shows the comparison between AISHE and AIOHE.

Fig. 5(a) shows the FMR curves of the YIG/Pt(2)/Fe(12) sample which showed a reduction in the linewidth at the Fe resonance when compared to the YIG/Fe(12) sample. Fig. 5(b) shows the SP-FMR signals as a function of the plane for the three samples. We note a strong anisotropy in the $\alpha = 60°$ sample and also a considerable value for the sample with $\alpha = 30°$. The sample deposited at $\alpha = 0°$ did not show any anisotropy. Fig. 5(c), (d) and (e) show the impressive result. For the first sample, no AISHE or AIOHE signal is observed (Fig. 5(c)). For the sample with intermediate anisotropy, a signal with a magnitude of approximately 25 nA appears (Fig. 5(d)), associated with AISHE and AIOHE in Fe$^{60}$(12). For the sample with strong anisotropy (Fig. 5(e)), the signal magnitude increases to 40 nA, which is more intense than the pure AISHE in Fe$^{60}$(12) shown in Fig. 3(e). The inset of Fig. 5(e) shows a comparison between the SP-FMR signal for YIG/Fe$^{60}$(12) and YIG/Pt(2)/Fe$^{60}$(12).

## III. CONCLUSION

In conclusion, we presented results on spin and orbital dynamics, highlighting their interplay in advanced magnetic systems. YIG/Fe samples were fabricated using the oblique DC sputtering deposition technique with an applied magnetic field, leading to the emergence of strong uniaxial anisotropy. This anisotropy serves as a magnetic order parameter. SP-FMR measurements revealed very small ISHE signals compared to the SP-FMR signal observed in YIG/Pt(2). The control of the magnetic order parameter enabled the emergence of the SP-FMR signal in the out-of-plane configuration, where the spin current's polarization and direction are collinear.

Orbital dynamics were investigated using samples with a Pt(2) spacer layer. YIG/Pt($t_{Pt}$)/Fe(12) samples were fabricated using the same deposition method. In in-plane SP-FMR measurements, for $t_{Pt} = 2$ nm, we observed a 3.5-fold signal enhancement compared to $t_{Pt} = 0$ nm, reflecting an enhanced signal due to IOHE in Fe(12). For out-of-plane measurements, on samples without magnetic anisotropy, we did not observe any SP-FMR signal. However, when strong anisotropy was induced, we observed a signal in YIG/Pt(2)/Fe(12) with a magnitude 4.0-fold higher than that in YIG/Fe(12), which we attributed to the orbital-to-charge conversion by AIOHE in the Fe(12) film.

These groundbreaking findings provide significant insights into the intricate interactions between spin and orbital currents. This understanding is expected to contribute to the development of next-generation orbitronic devices, paving the way for new applications in spintronic and orbitronic technologies.

## ACKNOWLEDGMENTS

This research was supported by the Conselho Nacional de Desenvolvimento Científico e Tecnológico (CNPq), Coordenação de Aperfeiçoamento de Pessoal de Nível Superior (CAPES) (Grant No. 0041/2022), Financiadora de Estudos e Projetos (FINEP), Fundação de Amparo à Ciência e Tecnologia do Estado de Pernambuco (FACEPE), Universidade Federal de Pernambuco, Multiuser Laboratory Facilities of DF-UFPE, Fundação de Amparo à Pesquisa do Estado de Minas Gerais (FAPEMIG) - Rede de Pesquisa em Materiais 2D e Rede de Nanomagnetismo, and the INCT of Spintronics and Advanced Magnetic Nanostructures (INCT SpinNanoMag), CNPq 406836/2022-1.

## DATA AVAILABILITY STATEMENT

The data supporting the findings of this study are available from the corresponding author upon reasonable request.

**Appendix A:** parameters obtained from theoretical fits

| $\alpha$ (°) | $4\pi M_{eff}$ (G) | $H_u$(Oe) | $\phi_u$(°) | $\gamma$ (CGS) |
|---|---|---|---|---|
| 60 | 16.0 | 571.0 | 90.8 | 2.7 |

**Table 1.** In plane FMR measurements for YIG/Fe(12), where $0° \leq \phi_H \leq 360°$

| $\alpha$ (°) | $4\pi M_{eff}$ (G) | $H_u$(Oe) | $\phi_u$(°) | $\gamma$ (CGS) |
|---|---|---|---|---|
| 30 | 18.8 | 77.2 | 132.5 | 2.84 |
| 60 | 16.57 | 500.4 | 93.0 | 2.7 |

**Table 2.** In plane FMR measurements for YIG/Pt(2)/Fe(12), where $0° \leq \phi_H \leq 360°$.

| substrate | $M_s$(G) | $H_d$(Oe) | $H_c$(Oe) | $\gamma$ (CGS) |
|---|---|---|---|---|
| Fe | $15 \times 10^3$ | -482.0 | -10.0 | 2.75 |
| YIG | 143 | - | -156.0 | 2.8 |

**Table 3.** out-of-plane FMR measurements for YIG and Fe, where $0° \leq \theta_H \leq 90°$ for YIG, and $10° \leq \theta_H \leq 90°$ for Fe films.